\documentstyle[epsfig,preprint,eqsecnum,aps,pre,amssymb]{revtex}

\begin{document}
\draft
\title{Thermoconvection in magnetized ferrofluids:
The influence of boundaries with finite heat conductivity }
\author{A.~Recktenwald and M.~L\"{u}cke\\}
\address{Institut f\"{u}r Theoretische Physik, Universit\"{a}t 
des Saarlandes,
         D-66041~Saarbr\"{u}cken,\\
         Federal Republic of Germany\\
}


\maketitle

\begin{abstract}
Realistic boundaries of finite heat conductivity 
for thermoconvection in a 
Rayleigh-B\'enard setup with magnetized 
ferrofluids are investigated. 
A linear stability analysis of the conductive state is performed with a 
shooting method. It shows that the critical wave number is for any magnetic 
field stronly influenced by the conductivity of the boundaries.
Linear as well as nonlinear coefficients of a Ginzburg Landau 
amplitude equation for convection shortly above the onset are evaluated as
functions of the magnetic Rayleigh number, the boundary conductivities, and 
the fluid Prandtl number.

\end{abstract}


%
%
\narrowtext

\newcommand{\bvec}[1]{{\bf #1}}

\section{Introduction}

Convection of magnetic fluids  
in the classical Rayleigh-B\'enard setup in the presence of a 
vertical magnetic field  has been studied 
theoretically by Finlayson \cite{F70}. 
However, the experiments  by Schwab et al. \cite{SHS83,Sc89} showed that 
the theoretical critical wave numbers of the flow structure were larger than 
those that were measured in the experiments done with horizontal plates of 
finite heat conductivity. In this paper we investigate how with 
decreasing conductivity of the plates linear and weakly nonlinear convection
properties change.

We consider a horizontal fluid layer of height $d$ between  
horizontal plates each of thickness $d_P$ 
subjected to a 
homogeneous vertical (z-direction) gravitational acceleration $g$ and 
a homogeneous vertical magnetic field. 
The ratio of the termal conductivities of the plates, $\Lambda_P$, and of 
the fluid, $\Lambda_F$, is $\zeta=\frac{\Lambda_P}{\Lambda_F}$ and 
$\delta=\frac{d_P}{d/2}$ is the plate thickness in
units of half the height of the fluid layer. 
Then the vertical temperature profile in the conductive, quiescent state 
without
convection is piecewise linear as shown in Fig.~\ref{fig_bildrb_warm_kond} for
the case of $\delta= 1$. 
A temperature difference $\Delta T$ imposed between the outer 
surfaces of the plates at $z=\pm (\frac{d}{2}+d_P)$ implies a temperature difference 
\begin{equation}
\label{Delta T_F_def}
\Delta T_F=\frac{\Delta T}{1+\delta/\zeta}
\end{equation}
across the fluid layer in the conductive state. 
We use this conductive $\Delta T_F$ to 
define the Rayleigh number 
\begin{equation}
Ra= \frac{\alpha g\,  \,d^3 \Delta T_F}{\kappa \nu}   \label{ra_def}
\end{equation}
and the magnetic Rayleigh number 
\begin{equation}
N =\frac{\mu_0 K^2  d^2 (\Delta T_F)^2}{\kappa \eta (1+\chi)}  \label{n_def}
\end{equation}
as control parameters. A negative 
value of $\Delta T$ or of $\Delta T_F$ stands for heating of the upper plate 
and a positive one  for lower plate heating. 
Here $\alpha$ is the thermal expansion coefficent, $\kappa$ the 
thermal diffusivity, $\nu$ the kinematic viscosity, and $\eta$ the 
effective viscosity of the fluid. Furthermore,  
$\mu_0$ is the magnetic field constant, $\chi$ the magnetic susceptibility, and
$K$ is the pyromagnetic coefficient, which is almost proportional to 
the external magnetic field.

For convenience $Ra$ and $N$ are normalized by the 
critical Rayleigh number $Ra_c^0$ for onset of convection in a  
fluid bounded by plates of conductivity $\Lambda_P$ and thickness $d_P$
in the absence of a  magnetic field. So we introduce
\begin{equation}
r=\frac{Ra}{Ra_c^0},  \qquad n=\frac{N}{3 Ra_c^0}.
\end{equation}

\section{Equations and Boundary conditions} \label{equation}

Henceforth we scale positions by the height of the fluid layer $d$,
time by $\frac{d^2}{\kappa}$, velocity by $\frac{\kappa}{d}$, 
temperature by  $\Delta T_F$,  and the vertical
magnetic field by $\frac{d^2 K \beta}{1+\chi}$. Using this scaling we get
from the basic hydrodynamic field equations \cite{F70} the
following system of partial differential equations  
\begin{mathletters}
\label{pdes}
\begin{eqnarray}
\frac{1}{\sigma} \partial_t \nabla^2 w 
&=& \nabla^4 w +  (\partial_x^2 + \partial_y^2) 
       \left[(Ra +N)\theta - N \varphi \right] 
     - \frac{1}{\sigma} \bvec{\nabla} \times \bvec{\nabla} \times
    \left[(    \bvec{v} \cdot \bvec{\nabla}) \bvec{v}\right]_z \nonumber\\
& & + N \left[ \partial_y(\partial_y \theta \partial_z \varphi 
    - \partial_z \theta
      \partial_y \varphi ) + \partial_x (\partial_x \theta \partial_z \varphi
      - \partial_z \theta \partial_x \varphi) \right] \label{gl_nse_rotw} \\
\partial_t \theta
&=& \nabla^2 \theta + w-(\bvec{v} \cdot \bvec{\nabla})\theta \\
\nabla^2 \varphi
&=& \partial_z^2 \theta  \label{gl_nse_rotphi}
\end{eqnarray}
\end{mathletters}
for the deviations  from the quiescent state of heat conduction in the fluid. 
Here $w=\bvec{v}\cdot \bvec{e_z}$ is the vertical velocity field, $\theta=T-T_{cond}$, 
$\varphi=(\bvec{H}-\bvec{H_{cond}})\cdot \bvec{e_z}$, and 
$\sigma=\frac{\nu}{\kappa}$ is the Prandtl number.
The pressure was eliminated by applying twice the rotation to the momentum 
balance equation. 

The equations are solved subject to the following realistic boundary conditions:
The plates are rigid, thus enforcing the no-slip condition
\begin{mathletters}
\label{bcs}
\begin{equation}
\label{gl_rb_geschwindigkeit}
w=\partial_z w = 0 \qquad \mbox{at} \qquad z = \pm \frac{1}{2}
\end{equation}
 on the velocity field.
The boundary conditions on the deviations of the temperature in 
the plates, $\tilde{\theta}$,   and in the fluid, $\theta$, from the 
conductive profile of Fig.~\ref{fig_bildrb_warm_kond} are
\begin{eqnarray}
\partial_z \theta&=&\zeta\, \partial_z \tilde{\theta}\mbox{,}\qquad
\theta=\tilde{\theta}\label{gl_rb_temperatur}\qquad \mbox{at}
\qquad z=\pm \frac{1}{2}
\end{eqnarray}
to ensure continuity of the heat flux and of the temperature. With 
$\Delta T$ being imposed externally one has 
\begin{eqnarray}
  \tilde{\theta}=0 \qquad\qquad \mbox{ at }
                 z = \pm \frac{1}{2}(1+\delta)   \mbox{.} \label{gl_thetatilde}
\end{eqnarray}
The vertical magnetic field deviation from the magnetic field in the
conductive state, $\varphi=(\bvec{H}-\bvec{H_{cond}}) \cdot \bvec{e_z}$,
has to fulfill the condition  
\begin{equation}
(1+\chi) \partial_z \varphi \pm k \varphi =0  \qquad \mbox{at}\qquad 
z=\pm \frac{1}{2}\mbox{.}
\label{gl_rb_phi}
\end{equation}
\end{mathletters}
Here $k$ is the wave number of the convective roll structure that bifurcates
out of the conductive state. 
\section{Linear Analysis}  

We have performed a standard linear stability analysis of the conductive 
state for plane wave perturbation fields obeying the above boundary 
conditions (\ref{bcs}).
To that end  the linearized version of the field equations of the ferrofluid
(\ref{pdes}) coupled to the heat diffusion equation in the plates 
were solved by a shooting method \cite{SB78}. 
The resulting critical eigenfunctions are plotted in 
Fig.~\ref{lambeigenftuwzeta5} for 
different magnetic fields at $\delta=1$ and $\zeta=5$. The latter being the
conductivity ratio realized in the experimental setup of 
Schwab et al. \cite{SHS83,Sc89}.
The maximal amplitude of the vertical velocity $\hat{w}(z)$ increases with 
growing magnetic field, i.e., with growing $n$ while that of the 
lateral velocity $\hat{u}(z)$, of the temperature $\hat{\theta}(z)$, and 
of the magnetic field $\hat{\varphi}(z)$ is decreasing with increasing 
$n$ without much change in the profiles. The discontinuous slope of 
$\hat{\theta}(z)$ at $z =\pm \frac{1}{2}$ comes from the 
continuity of the heatflux. Here we used the normalization 
$\int dz\, \hat{w}(z)\, \hat{\theta}(z)=1$.

From the neutral stability problem one finds that the instability is a 
stationary one as for boundaries with infinite conductivity, $\zeta=\infty$. 
The critical Rayleigh number $r_c(n)$ and the corresponding 
critical wave number $k_c(n)$ as function of the magnetic 
Rayleigh number $n$ are shown in Figs.~\ref{lambdarakrit} and \ref{lambdakkrit},
respectively, for $\delta=1$ and several $\zeta$. Both, $k_c$ as well as $r_c$ shift to 
smaller values when the conductivity of the plates decreases. The variation of
$r_c(n)$ with $n$ can be fitted reasonably well by a 
linear law. The fit coefficients are presented in Table \ref{tab_warmraq}. 
for several values of  $\zeta$. 

The symbols in Figs.~\ref{lambdarakrit} and 
\ref{lambdakkrit} refer to Rayleigh numbers and wave numbers that were observed
at onset of convection in experiments done with fluid layers of different 
heights $d$ betweeen $1$ and $4$ mm and an upper plate of fixed thickness
$d_P=6$ mm with reduced conductivity $\zeta=5$ \cite{Sc89}. Our linear analysis showed 
for such values of $\delta \geq 3$ that $r_c$ and $k_c$ had well reached their
large-$\delta$ asymptotic plateau behaviour: Both, $r_c$ and $k_c$ decrease in
the range $0 < \delta < 1$ with growing $\delta$ and have approached
their large-$\delta$ constant asymptote around  $\delta= 1$ so that at 
$\delta > 1$ there is no more significant variation with $\delta$. Note in 
addition that the influence of $\delta$ on convection properties decreases with
growing conductivity $\zeta$ of the plate --- the thermal stress on the fluid,
$\Delta T_F$, that determines the driving (\ref{ra_def}) and (\ref{n_def}) 
approaches $\Delta T$ for
$\zeta \rightarrow \infty$ and the fluid approaches the limiting situation of
beeing bounded by perfectly heat conducting plates.

Our results agree in  the nonmagnetic limit, $n=0$, with the linear stability 
analysis of Jenkins et al. \cite{JP84} for all $\zeta$.   
Furthermore, in the presense of a magnetic field we confirm the theoretical 
results of Schwab \cite{Sc89} and Stiles et al. \cite{SK90II} reported for the 
limit $\zeta=\infty$. We have no explanation for the fact that the wave numbers
(symbols in Fig.~\ref{lambdakkrit}) that were observed to be selected by
convective 
patterns close to onset in the experimental setup of Schwab et al. \cite{Sc89}
lie closer to the $\zeta=1$ theoretical critical wave numbers than to the 
$\zeta=5$ curve.

\section{Weakly Nonlinear analysis}  \label{amplitudewarm}

Shortly above the stability threshold the nonlinear convective state can 
approximately be described by
\begin{equation}
\vec{Y}(x,z,t)=(u,w,\theta,\varphi)^T= 
A(x,t)\, \vec{\hat{Y}}(z)\, e^{i\,k_c\, x} + c.c. \mbox{.}
\label{Amp}
\end{equation}
Here $\vec{\hat{Y}}(z)$ is the  eigenvector of the linear problem at the
critical point presented in Fig.~\ref{lambeigenftuwzeta5} and
$A(x,t)$ is the saturation amplitude of the critical mode determined by the 
nonlinearity in the 
field equations. For slightly supercritical driving 
($0 \le \epsilon=r-r_c \ll 1$) 
a  small band of lateral wave numbers with 
$|k-k_c| =  {\cal O}(\epsilon^{\frac{1}{2}})$ is 
excited giving  rise to a slow spatiotemporal variation  of the 
convective roll structure.
These slow  variations of the amplitude $A$ are conveniently captured by 
the  method of multiple scales \cite{BO78}. 
It allows to derive in a straightforward manner the 
Ginzburg-Landau amplitude equation  \cite{S69} 
\begin{equation}
\tau_0\partial_t A(x,t)=\left[ \epsilon+\xi_0^2 \partial_x^2
   -\gamma|A(x,t)|^2\right] A(x,t) 
\end{equation}
from the hydrodynamic field equations. 

We have determined the coefficients $\tau_0$, $\xi_0^2$, and $\gamma$ of this
equation for rigid boundaries with different $\zeta$ by numerically 
evaluating special scalar products of the eigenfunctions 
\cite{RLM93}.
Since the coefficients with no magnetic field, i.e., for an ordinary 
fluid have not yet been presented in the literature as a function of 
$\zeta$ we show them in Fig.~\ref{fig_gamma_warm_no0} for $\delta=1$. 
The linear coefficients $\tau_0$ and $\xi_0^2$ are presented in
Fig.~\ref{tauxiwarmno} as  functions of the magnetic Rayleigh number $n$.
The nonlinear coefficient $\gamma$ shows a linear dependence on $n$, i.e.,
\begin{equation} 
\gamma(\zeta,n)=\gamma_0(\zeta) + b_\gamma(\zeta) \, n \, .
\label{gamma_zeta_n}
\end{equation}
The expansion coefficients $\gamma_0(\zeta)$ and $b_\gamma(\zeta)$ are given 
in Table~\ref{tab_gamma_warm} for several $\zeta$. 

To evaluate the slope $\frac{\partial Nu}{\partial n}$ of the 
Nusselt number at the critical point we start from the relation
\begin{equation}
Nu(r,n) = 1 + \frac{r-r_c(n)}{\gamma(n)}
\end{equation}
that holds close to onset, $r-r_c(n) \ll 1$. Thus
\begin{equation}
\left.\frac{\partial Nu}{\partial n}\right|_c =
-\frac{1}{\gamma(n)} \frac{\partial r_c(n)}{\partial n} \, .
\label{dnudnc}
\end{equation}
Since $r_c(n) \simeq 1 - b_r\, n$ one can rewrite (\ref{gamma_zeta_n}) as 
$\gamma = \gamma_0 + b_\gamma(1-r_c)/b_r$ to obtain
$\left.\frac{\partial Nu}{\partial n}\right|_c$ in the more compact form
\begin{equation}
\left.\frac{\partial Nu}{\partial n}\right|_c =
\frac{1}{ a - b \,r_c} \mbox{.}
\label{gl_dnudnfitnowarm}
\end{equation}
The fit coefficients $b=\frac{b_\gamma}{b_r^2}$
and $a=b + \frac{\gamma_0}{b_r}$ are presented in Table \ref{tab_dnudnwarm}. 

\section{Prandtl number dependence}  \label{Prandtldepend}

The results 
presented so far refer to fluids of Prandtl number $\sigma = 1$. Now we 
discuss the $\sigma $ dependence for the case of perfectly heat conducting
plates, $\zeta=\infty$. The $\sigma $ dependence for finite $\zeta$, say,
$\zeta \gtrsim 1$ is similar.
The marginal stability curve and the linear eigenfunctions of the stationary
eigenvalue problem have no Prandtl number dependence. Thus  
$r_c$, $k_c$, and $\xi_0^2$ are $\sigma$-independent. 
But $\tau_0$ depends on $\sigma$. 
Graphs for different  $n$ are given in Fig.~\ref{fig_tau0sigma} for
$\zeta=\infty$. Note that the difference 
$\tau_0(\sigma,n) - \tau_0(\sigma=1,n)$ is practically independent of $n$.
This implies that 
$\tau_0(\sigma,n)$ can be approximated well by 
\begin{equation}
\tau_0(\sigma,n) - \tau_0(\sigma=1,n) =  \frac{\sigma + 0.5117}{19.65\, \sigma}
                               - 0.0769
\label{gl_tau0noapprox}
\end{equation}
for $\zeta=\infty$ and the right hand side of (\ref{gl_tau0noapprox})
is just $\tau_0(\sigma,n=0)$ \cite{CH93} for the case without magnetic field.
Thus one only needs to know $\tau_0(\sigma=1,n)$ which can be taken from the 
long-dashed curve in the lower part of Fig.~\ref{fig_tau0sigma}. It represents
$\tau_0(\sigma=1,n,\zeta=1000)$ which is practically the same as 
$\tau_0(\sigma=1,n,\zeta=\infty)$.

The $\sigma$ dependence of the nonlinear coefficient $\gamma$ can be fitted by
the expression 
\begin{equation}
\gamma(n,\sigma) = \gamma^{(0)}(n)  
                  -\frac{\gamma^{(1)}(n)}{\sigma}
                  +\frac{\gamma^{(2)}(n)}{\sigma^2}\mbox{.}
\label{gl_gafit}
\end{equation}
The fit coefficients $\gamma^{(i)}(n)$ are presented in
Table \ref{tab_gammasigmano} for several $n$.
For $\sigma=1$ the $n$ variation 
is practically linear, i.e., $\gamma(n,\sigma=1) = 0.7027 + 1.42427 \, n $. 
Graphs of the difference $\gamma(n,\sigma) - \gamma(n,\sigma=1)$ are given in
Fig.~\ref{fig_gamma-gamma0}. 
For Prandtl number larger than $10$ the dependence 
on $\sigma$ can be neglected and one can use $\sigma=\infty$.

Support by the Deutsche Forschungsgemeinschaft is gratefully acknowledged.

\begin{table}
\begin{center}
\begin{tabular}{|c|c|c|c|c|c|c|c|c|}\hline
$\zeta$ & $0.8$&  $1$ & $2$ & $5$ & $10$ & $100$ & $1000$ & $\infty$   \\ \hline
\tableline
\multicolumn{9}{|c|}{$r_c=1- b_r n$}\\
\tableline
$Ra_{c}^0$ & $1256.9669$&  $1303.4412$ & $1439.8249$ & $1574.0344$ & $1634.8154$ & $1699.8132$ & $1706.9597$ & $1707.7618$   \\ \hline
$b_r$ & $2.424$&  $2.403$ & $2.349$ & $2.304$ & $2.283$ & $2.265$ & $2.262$ & $2.259$   \\ \hline
\tableline
\multicolumn{9}{|c|}{$\gamma=\gamma_0+b_\gamma \,n$}\\
\tableline
$\gamma_0$ & $0.6120$&  $0.6145$ & $0.6330$ & $0.6634$ & $0.6802$ & $0.6991$ & $0.7023$ & $0.7027$   \\ \hline
$b_\gamma$ & $1.0709$&  $1.1223$ & $1.2656$ & $1.3883$ & $1.4468$ & $1.4992$ & $1.4594$ & $1.4243$   \\ \hline
\tableline
\multicolumn{9}{|c|}{$\left.\frac{\partial Nu}{\partial n}\right|_c=
\frac{1}{a - b \, r_c}$}\\
\tableline
$a$ & $0.434$&  $0.450$ & $0.498$ & $0.549$ & $0.575$ & $0.601$ & $0.596$ & $0.590$   \\ \hline
$b$ & $0.182$&  $0.194$ & $0.229$ & $0.262$ & $0.277$ & $0.292$ & $0.285$ & $0.279$   \\ \hline
\end{tabular}
\end{center}
\caption{Fit-coefficients for the $n$ dependence of $r_c, \gamma$, and
$\left.\frac{\partial Nu}{\partial n}\right|_c$ for different $\zeta$.}
\label{tab_warmraq}\label{tab_gamma_warm}\label{tab_dnudnwarm}
\end{table}

\begin{table}[th]
\begin{center}
\begin{tabular}{|r|ccc|}\hline
$n$ &$\gamma^{(0)}$ &$\gamma^{(2)}$ &$\gamma^{(1)}$ \\ \hline
0     & 0.699    &   0.00832  & 0.00471 \\
0.01  & 0.710    &   0.00966  & 0.00453 \\
0.04  & 0.745    &   0.00941  & 0.00547 \\
0.08  & 0.792    &   0.00916  & 0.00673 \\
0.2   & 0.945    &   0.00851  & 0.00102 \\
0.4   & 1.212    &   0.00728  & 0.01411  \\ 
0.8   & 1.756    &   0.00568  & 0.01867  \\ 
1     & 2.014    &   0.00502  & 0.02035  \\
\end{tabular}
\end{center}
\caption{Fit coefficients for the $\sigma$ dependence of 
$\gamma(n,\sigma) = \gamma^{(0)}(n)  
                  -\frac{\gamma^{(1)}(n)}{\sigma}
                  +\frac{\gamma^{(2)}(n)}{\sigma^2}$
for different $n$.}
\label{tab_gammasigmano}
\end{table}



\begin{figure}
\caption{
Vertical
 temperature profile of  the quiescent conductive state in the plates 
(P) and in the fluid (F) here shown for 
$\delta=\frac{d_P}{d/2}=1$. With $\Delta T$ beeing imposed externally the 
 conductive temperature difference across the fluid layer is 
$\Delta T_F=\frac{\Delta T}{1+\delta/\zeta}$.
}
\label{fig_bildrb_warm_kond}
\end{figure}

%
%
\begin{figure}
\caption{
Vertical variation of the linear critical 
eigenfunctions of lateral 
velocity $\hat{u}$, vertical velocity $\hat{w}$, 
magnetic field $\hat{\varphi}$, and 
temperature $\hat{\theta}$ for 
$\zeta=5, \delta=1$, and different  
reduced magnetic Rayleigh numbers $n$. 
}
\label{lambeigenftuwzeta5}
\label{lambeigenftthetazeta5}
\end{figure}

%
%
%
\begin{figure}
\caption{
Critical Rayleigh number $r_c$ versus magnetic Rayleigh number 
$n$ for different $\zeta$ and $\delta=1, \chi=1$. The curves for $\zeta=5$ and 
$\zeta=\infty$ are practically identical. Symbols refer to experimentally
observed convective thresholds [3]
in setups with $\zeta=5$ and
$\delta=$ 3($\triangledown$), 4($\circ$), 6($\bigtriangleup$),
8($\Box$), and 12($\diamond$). See text for further discussion.}
\label{lambdarakrit}
\end{figure}

%
%
%
\begin{figure}
\caption{
Critical wave number $k_c$ versus  magnetic Rayleigh number 
$n$ for different $\zeta$ and $\chi=1$. Symbols refer to experimentally
observed wave numbers of convection patterns close to threshold [3]
in setups with $\zeta=5$ and different $\delta $ identified in
 Fig.~\ref{lambdarakrit}. 
See text for further discussion.
}
\label{lambdakkrit}
\end{figure}

%
%
\begin{figure}
\caption{
Coefficients  of the amplitude equation as functions 
of $\zeta$ without magnetic field, $n = 0$. The normalization of the critical 
eigenfunction entering into $\gamma$ was chosen to be 
$\int dz\, \hat{w}(z)\, \hat{\theta}(z)=1$.
}
\label{fig_gamma_warm_no0}
\label{tauxiwarmno0}
\end{figure}

%
%
%
\begin{figure}
\caption{Linear coefficients $\tau_0$  and $\xi_0^2$ of the amplitude 
equation as functions of the magnetic Rayleigh number $n$ for 
different $\zeta$. The long-dashed curves for $\zeta=1000$ are practically 
the same as for perfectly heat conducting plates with $\zeta=\infty$.
}
\label{tauxiwarmno}
\end{figure}

%
%
%
\begin{figure}
\caption{
Initial slope of the Nusselt number versus critical Rayleigh number.
Symbols refer to experimental results obtained with plates of large heat 
conductivity [3].
}
\label{DnuDnofslopenowarm}
\end{figure}

\begin{figure}
\caption{Linear coefficient  $\tau_0(\sigma,n) - \tau_0(\sigma=1,n)$
versus $\sigma$ for different magnetic Rayleigh numbers $n$.}
\label{fig_tau0sigma}
\end{figure}
%

%
\begin{figure}
\caption{Nonlinear coefficient $\gamma(\sigma,n) - \gamma(\sigma=1,n)$ 
versus $\sigma$ for different magnetic Rayleigh numbers $n$.
}
\label{fig_gamma-gamma0}
\end{figure}

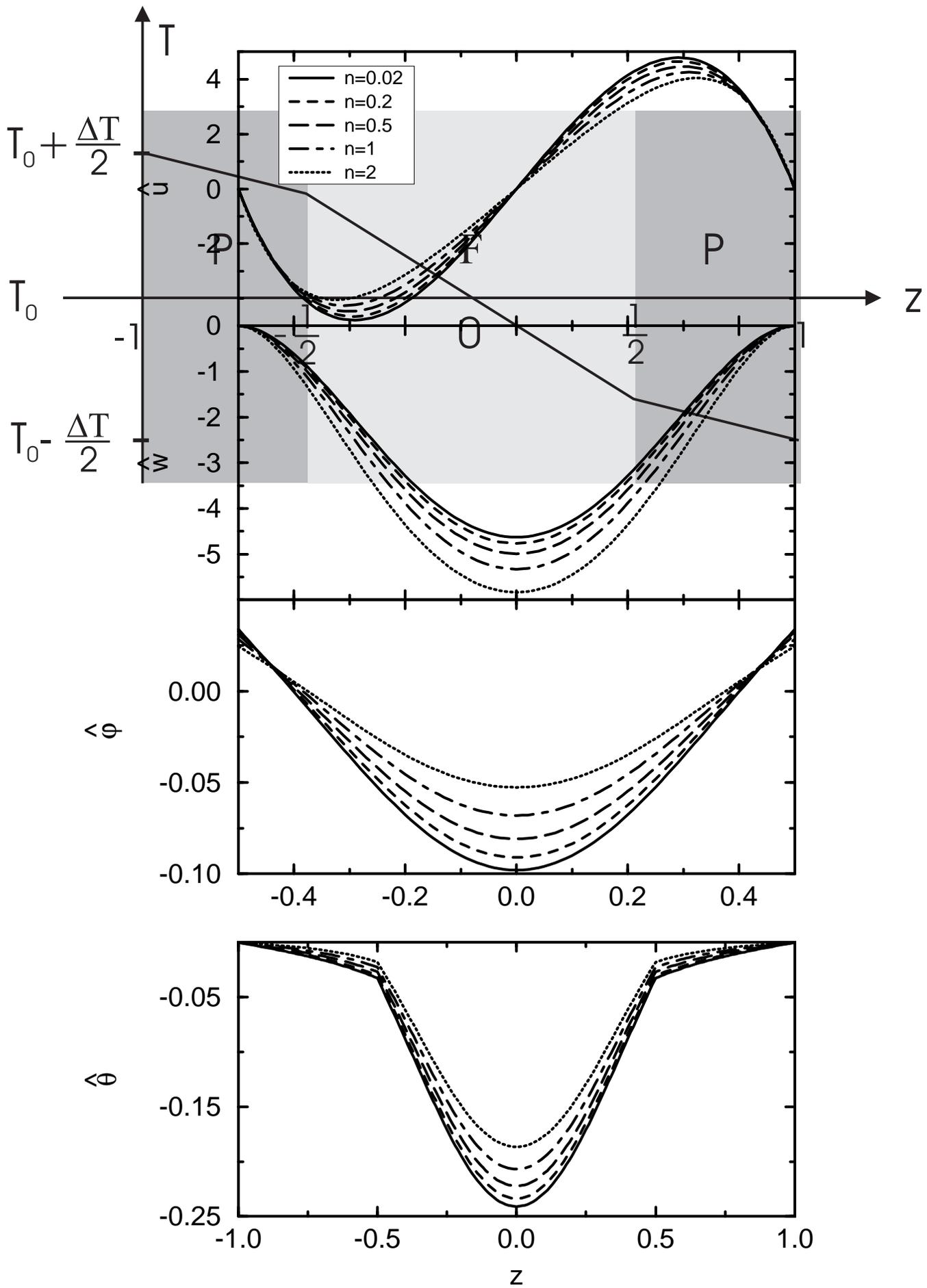
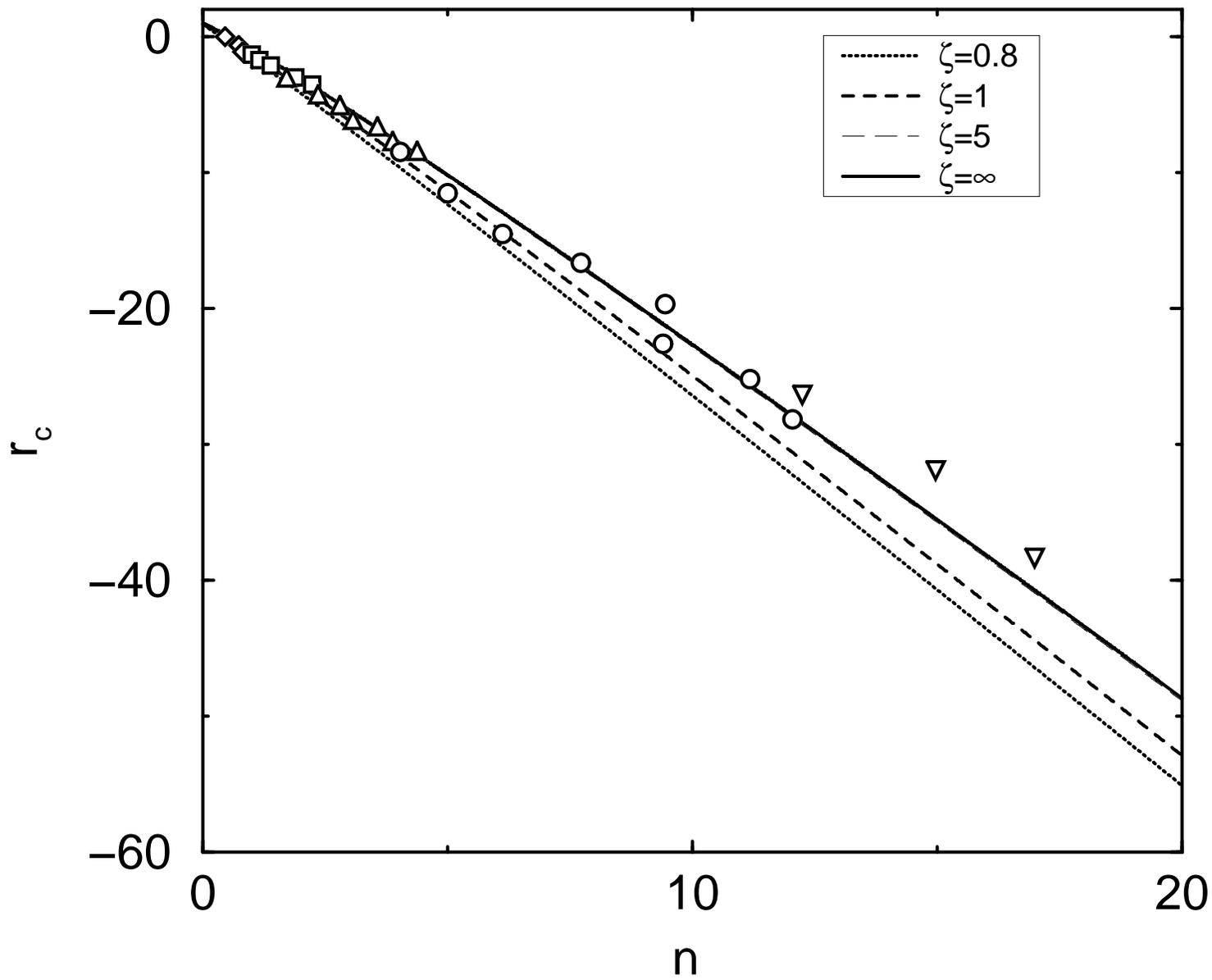
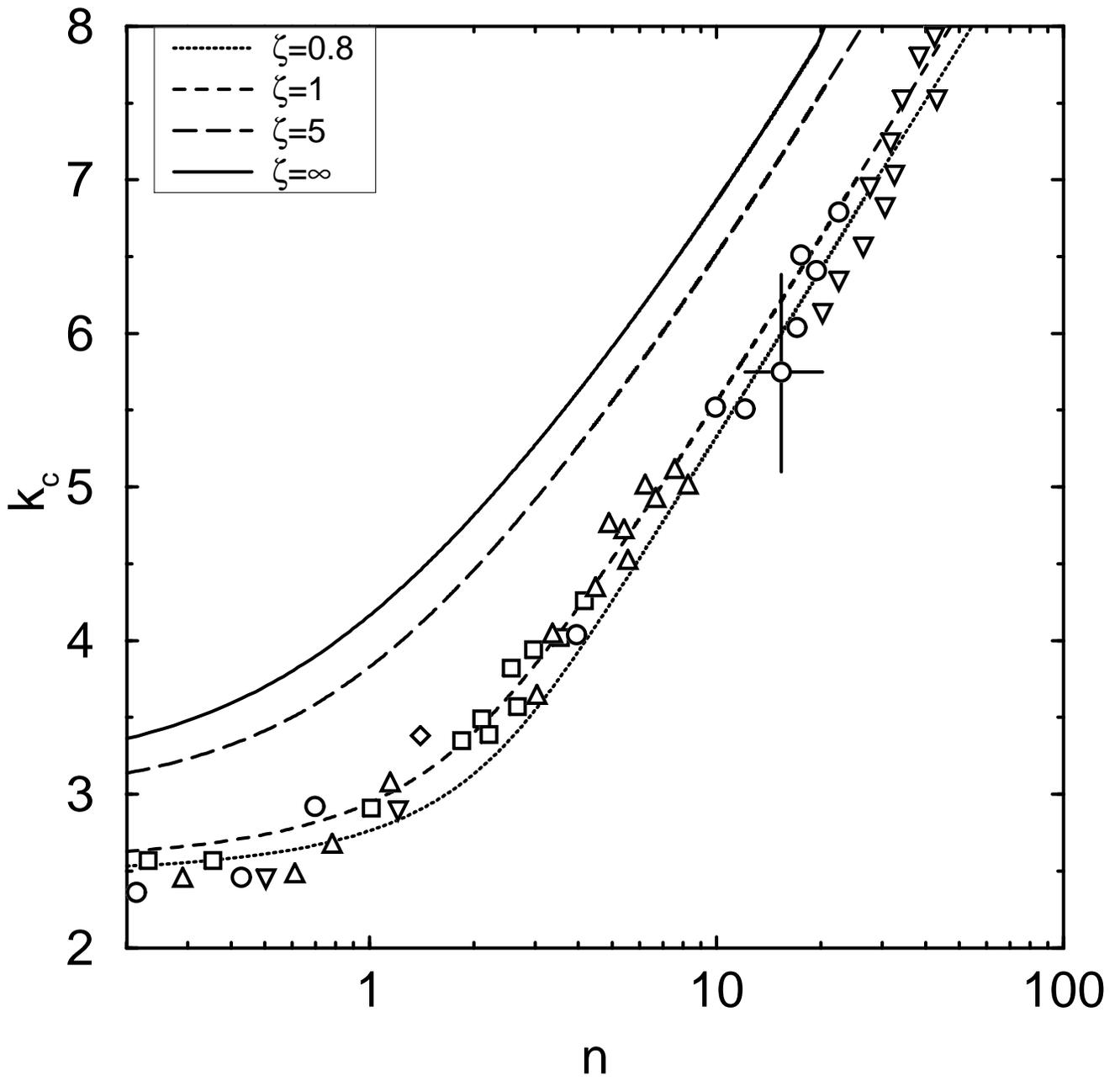
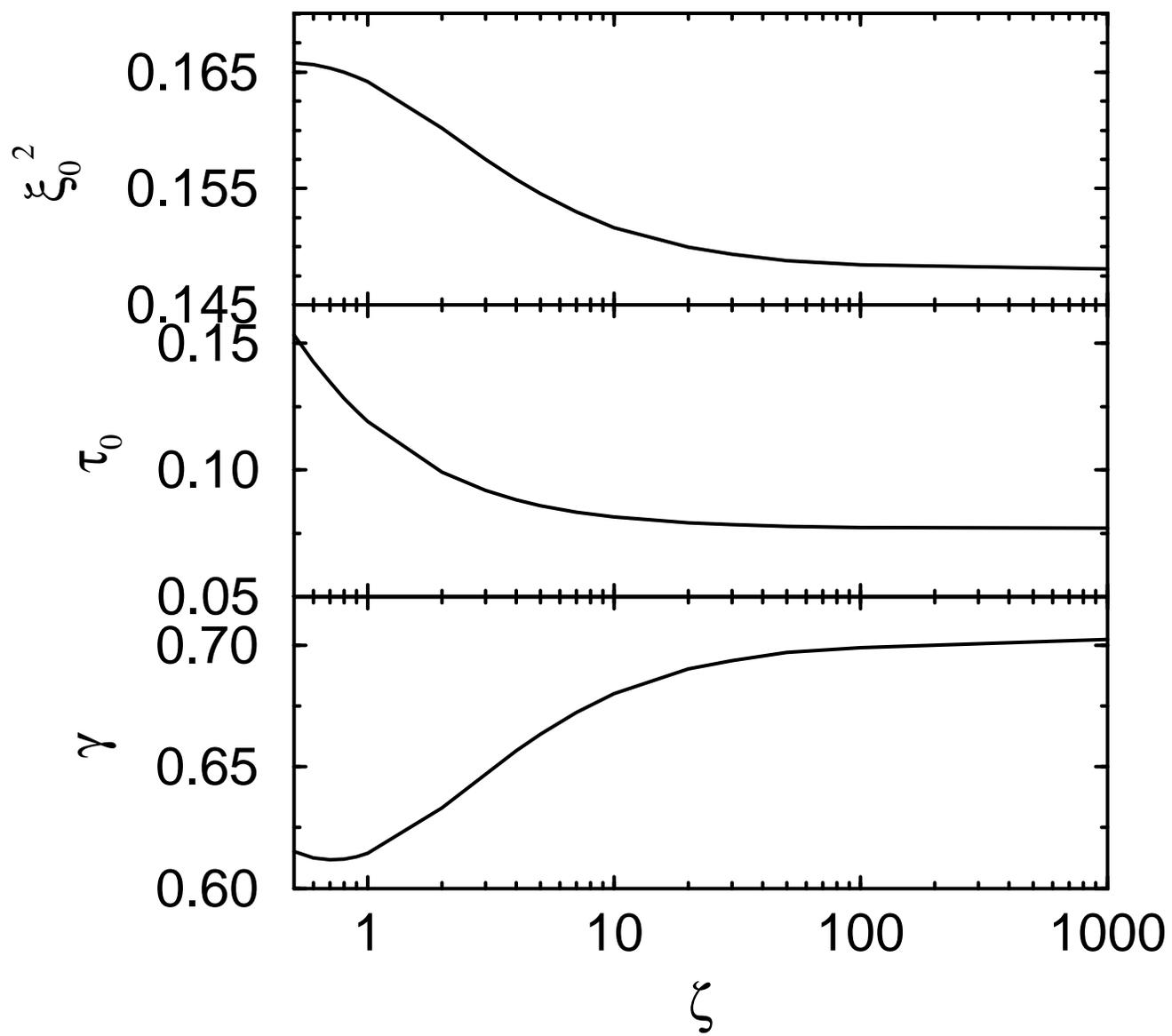
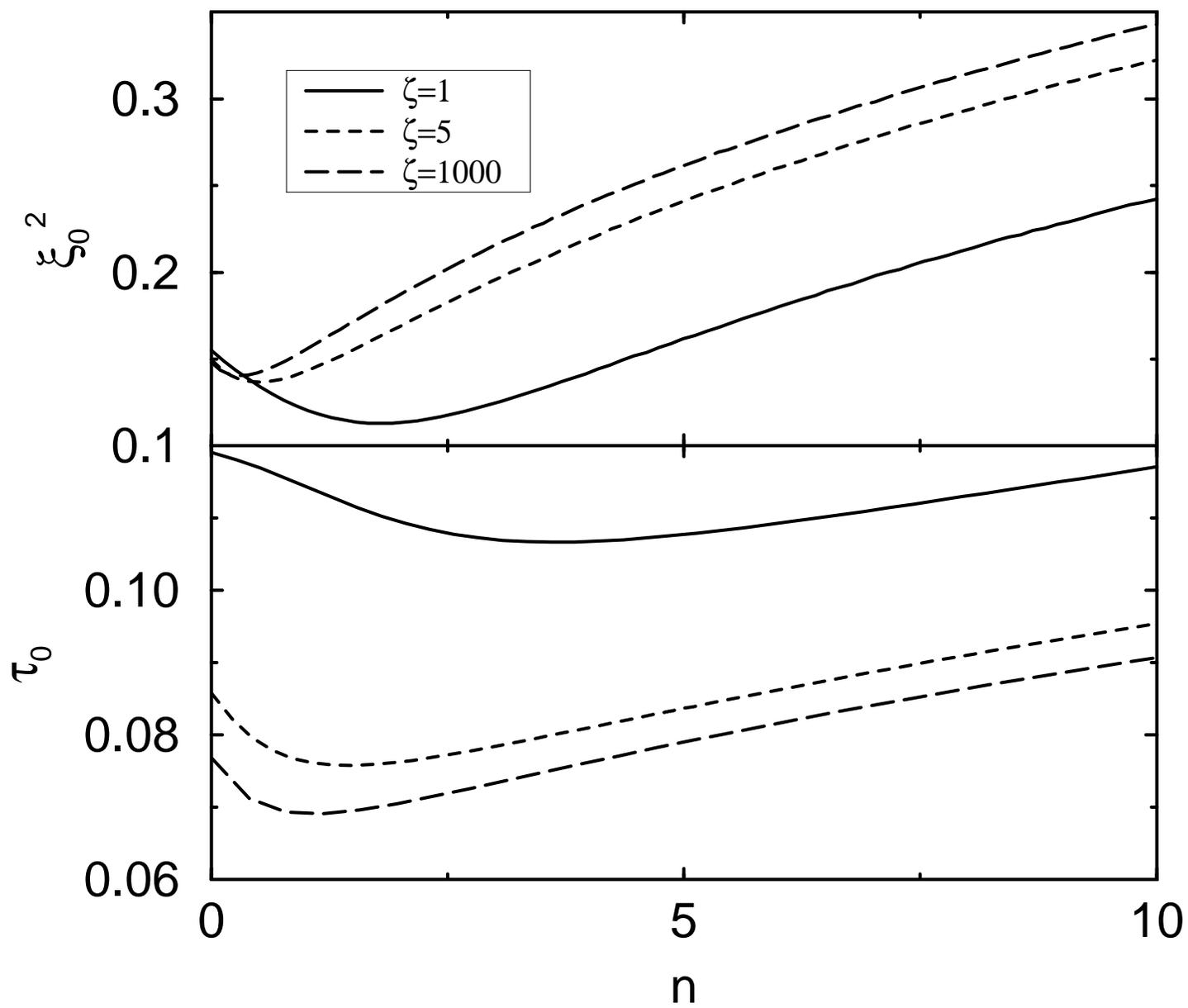
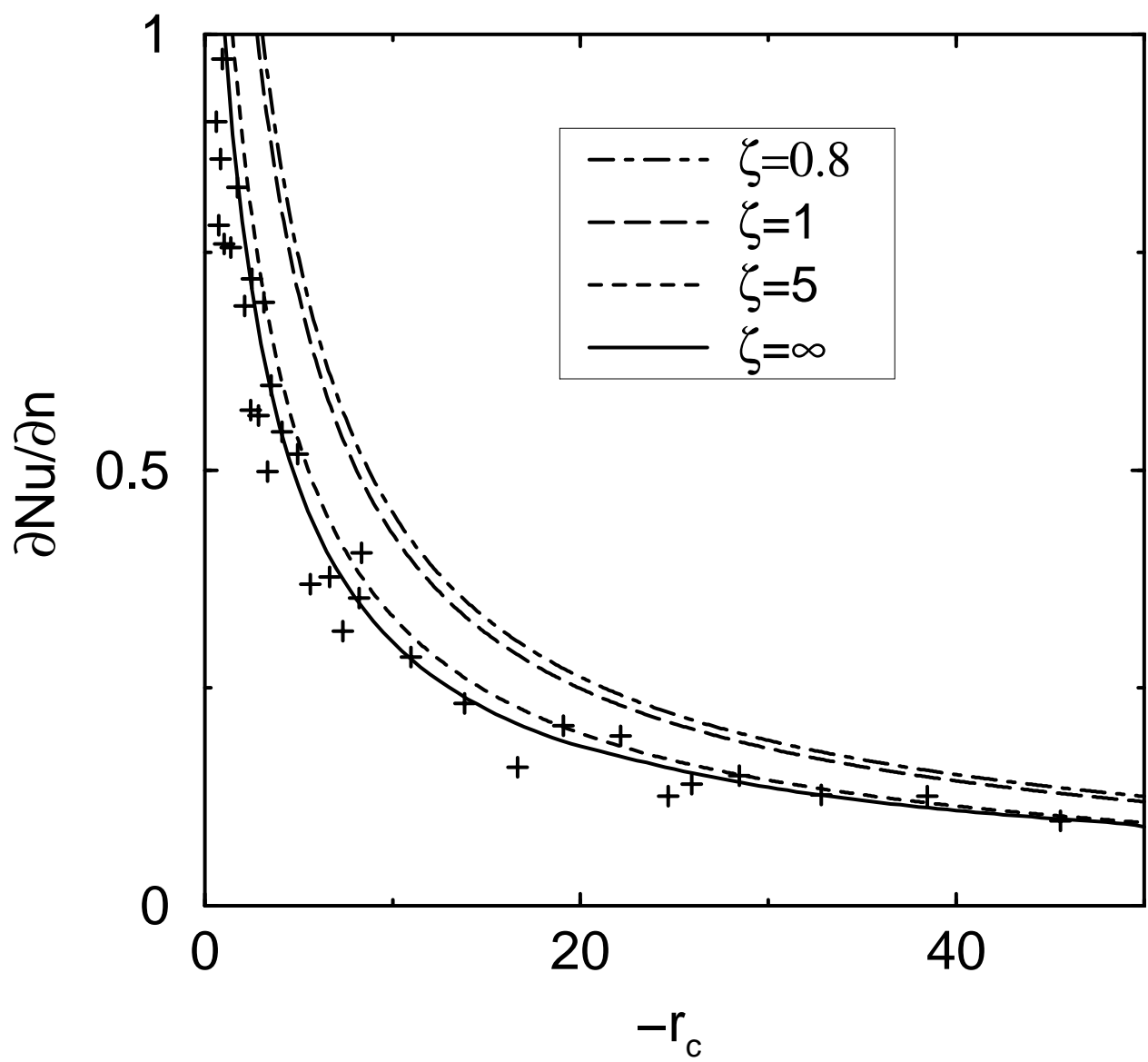
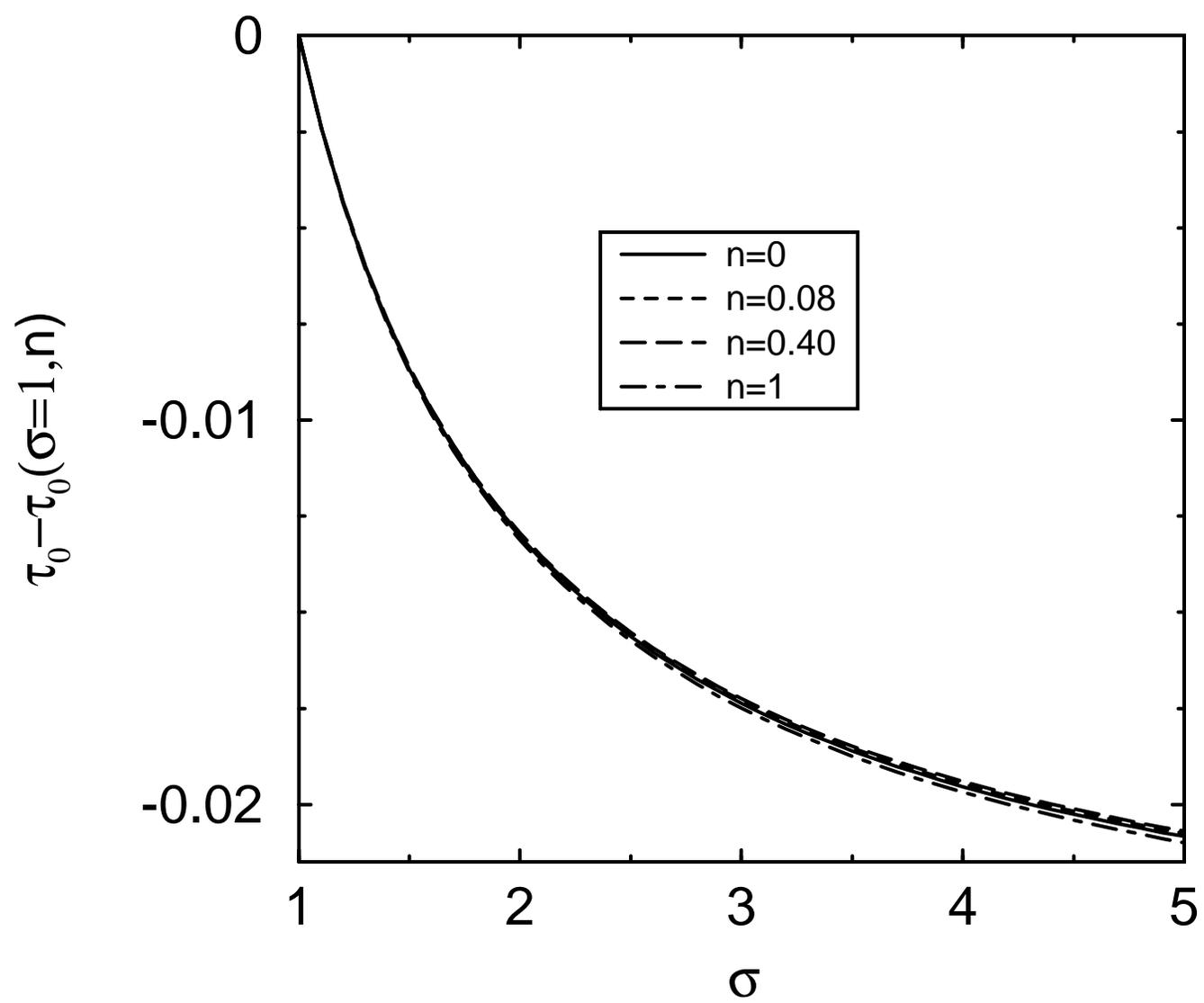
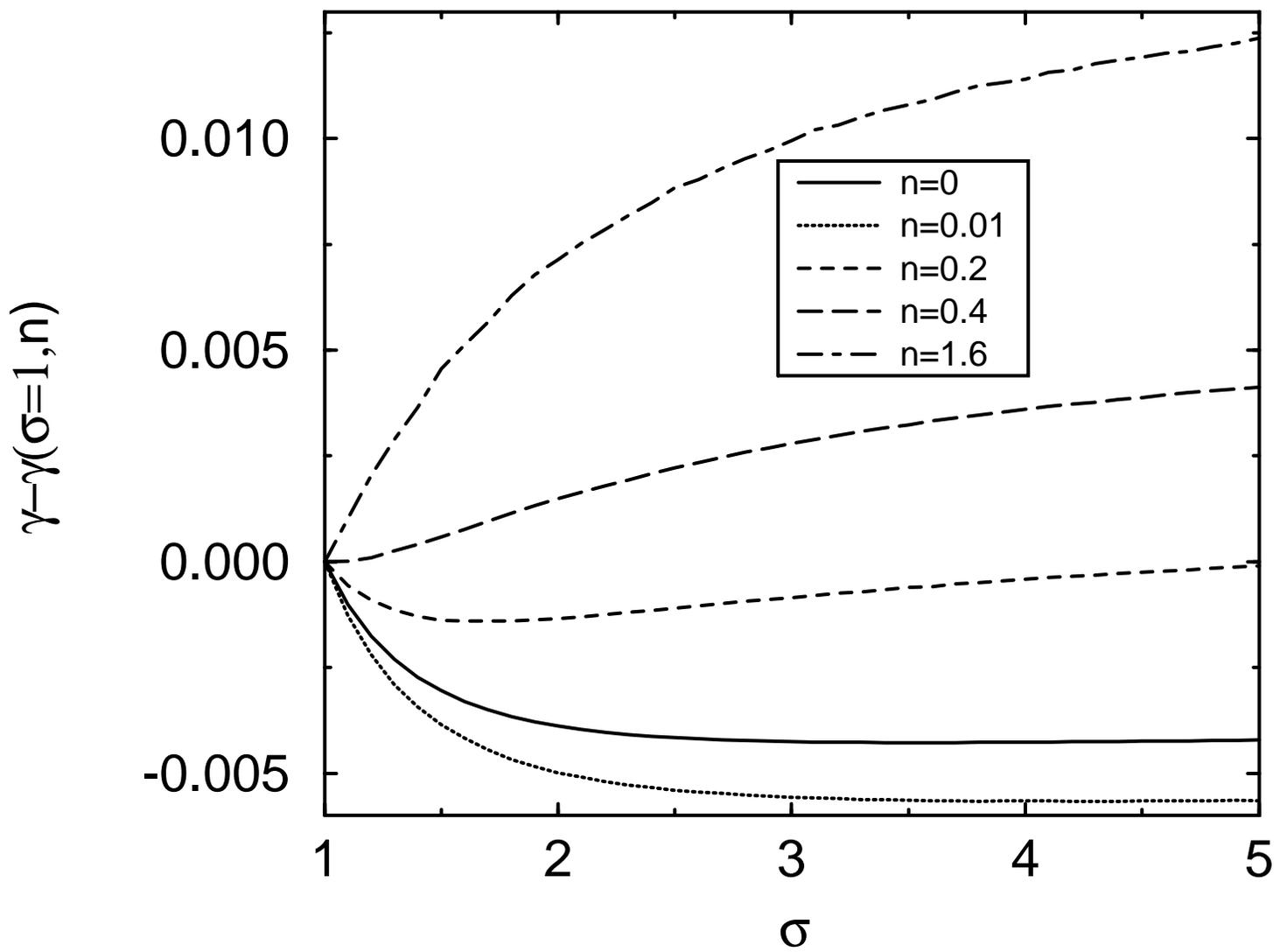
\end{document}